# On Flexible Web Services Composition Networks

Chantal Cherifi[1], Vincent Labatut[2], Jean-François Santucci[1]

[1] University of Corsica, UMR CNRS, SPE Laboratory, Corte, France
[2] Galatasaray University, Computer Science Department, Istanbul, Turkey
chantalbonner@gmail.com

**Abstract.** The semantic Web service community develops efforts to bring semantics to Web service descriptions and allow automatic discovery and composition. However, there is no widespread adoption of such descriptions yet, because semantically defining Web services is highly complicated and costly. As a result, production Web services still rely on syntactic descriptions, key-word based discovery and predefined compositions. Hence, more advanced research on syntactic Web services is still ongoing. In this work we build syntactic composition Web services networks with three well known similarity metrics, namely Levenshtein, Jaro and Jaro-Winkler. We perform a comparative study on the metrics performance by studying the topological properties of networks built from a test collection of real-world descriptions. It appears Jaro-Winkler finds more appropriate similarities and can be used at higher thresholds. For lower thresholds, the Jaro metric would be preferable because it detect less irrelevant relationships.

## 1 Introduction

Web Services (WS) are autonomous software components that can be published, discovered and invoked for remote use. For this purpose, their characteristics must be made publicly available under the form of WS descriptions. Such a description file is comparable to an interface defined in the context of object-oriented programming. It lists the operations implemented by the WS. Currently, production WS use syntactic descriptions expressed with the WS description language (WSDL) [1], which is a W3C (World Wide Web Consortium) specification. Such descriptions basically contain the names of the operations and their parameters names and data types. Additionally, some lower level information regarding the network access to the WS is present. WS were initially designed to interact with each other, in order to provide a composition of WS able to offer higher level functionalities. Current production discovery mechanisms support only keyword-based search in WS registries and no form of inference or approximate match can be performed.

    WS have rapidly emerged as important building blocks for business integration. With their explosive growth, the discovery and composition processes have become extremely important and challenging. Hence, advanced research comes from the semantic WS community, which develops a lot of efforts to bring semantics to WS

descriptions and to automate discovery and composition. Languages exist, such as OWL-S [2], to provide semantic unambiguous and computer-interpretable descriptions of WS. They rely on ontologies to support users and software agents to discover, invoke and compose WS with certain properties. However, there is no widespread adoption of such descriptions yet, because their definition is highly complicated and costly, for two major reasons. First, although some tools have been proposed for the annotation process, human intervention is still necessary. Second, the use of ontologies raises the problem of ontology mapping which although widely researched, is still not fully solved. To cope with this state of facts, research has also been pursued, in parallel, on syntactic WS discovery and composition.

Works on syntactic discovery relies on comparing structured data such as parameters types and names, or analyzing unstructured textual comments. Hence, in [3], the authors provide a set of similarity assessment methods. WS Properties described in WSDL are divided into four categories: lexical, attribute, interface and QoS. *Lexical similarity* concerns textual properties such as the WS name or owner. *Attribute similarity* estimates the similarity of properties with more supporting domain knowledge, like for instance, the property indicating the type of media stream a broadcast WS provides. *Interface similarity* focuses on the WS operations input and output parameters, and evaluates the similarity of their names and data types. *Qos similarity* assesses the similarity of the WS quality performance. A more recent trend consists in taking advantage of the latent semantics. In this context, a method was proposed to retrieve relevant WS based on keyword-based syntactical analysis, with semantic concepts extracted from WSDL files [4]. In the first step, a set of WS is retrieved with a keyword search and a subset is isolated by analyzing the syntactical correlations between the query and the WS descriptions. The second step captures the semantic concepts hidden behind the words in a query and the advertisements in the WS, and compares them.

Works on syntactic composition encompasses a body of research, including the use of networks to represent compositions within a set of WS. In [5], the input and output parameters names are compared to build the network. To that end, the authors use a strict matching (exact similarity), an approximate matching (cosine similarity) and a semantic matching (WordNet similarity). The goal is to study how approximate and semantic matching impact the network small-world and scale-free properties. In this work, we propose to use three well-known approximate string similarity metrics, as alternatives to build syntactic WS composition networks. Similarities between WS are computed on the parameters names. Given a set of WS descriptions, we build several networks for each metrics by making their threshold varying. Each network contains all the interactions between the WS that have been computed on the basis of the parameters similarities retrieved by the approximate matching. For each network we compute a set of topological properties. We then analyze their evolution for each metric, in function of the threshold value. This study enables us to assess which metric and which threshold are the most suitable.

Our main contribution is to propose a flexible way to build WS composition networks based on approximate matching functions. This approach allows to link some semantically related WS that does not appear on WS composition networks based on strict equality of the parameters names. We provide a thorough study regarding the use of syntactic approximate similarity metrics on WS networks topology. The results of our experimentations allow determining the suitability of the metrics and the threshold range that maintains the false positive rate at an acceptable level.

In section 2, we give some basic concepts regarding WS definition, description and composition. Interaction networks are introduced in section 3 along with the similarity metrics. Section 4 is dedicated to the network properties. In section 5 we present and discuss our experimental results. Finally, in section 6 we highlight the conclusions and limitations of, and explain how our work it can be extended.

## 2 Web Services

In this section we give a formal definition of WS, explain how it can be described syntactically, and define WS composition.

A WS is a set of operations. An operation i represents a specific functionality, described independently from its implementation for interoperability purposes. It can be characterized by its input and output parameters, noted $I_i$ and $O_i$, respectively. $I_i$ corresponds to the information required to invoke operation i, whereas $O_i$ is the information provided by this operation. At the WS level, the set of input and output parameters of a WS α are $I_\alpha = \cup I_i$ and $O_\alpha = \cup O_i$, respectively. Fig. 1 represents a WS labeled $\alpha$ with two operations numbered 1 and 2, and their sets of input and output parameters: $I_1 = \{a, b\}$, $O_1 = \{d\}$, $I_2 = \{c\}$, $O_2 = \{e, f\}$, $I_\alpha = \{a, b, c\}$, $O_\alpha = \{d, e, f\}$.

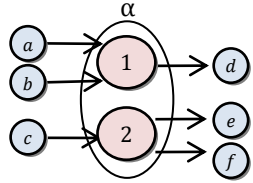

**Fig. 1.** Schematic representation of a WS $\alpha$, with two operations 1 and 2 and six parameters $a$, $b$, $c$, $d$, $e$ and $f$.

WS are either syntactically or semantically described. In this work, we are only concerned by the syntactic description of WS, which relies on the WSDL language. A WS is described by defining messages and operations under the form of an XML document. A message encapsulates the data elements of an operation. Each message consists in a set of input or output parameters. Each parameter has a name and a data type. The type is generally defined using the XML schema definition language (XSD), which makes it independent from any implementation.

WS composition addresses the situation when a request cannot be satisfied by any available single atomic WS. In this case, it might be possible to fulfill the request by combining some of the available WS, resulting in a so-called *composite* WS. Given a request $r$ with input parameters $I_r$, desired output parameters $O_r$ and a set of available WS, one needs to find a WS $\alpha$ such that $I_r \supseteq I_{w\alpha}$ and $O_r \subseteq O_\alpha$. Finding a WS $\alpha$ that can fulfill $r$ alone is referred to as WS discovery. When it is impossible for a single WS to fully satisfy $r$, one needs to compose several WS $\{\alpha, \beta, ..., \eta\}$, so that for all $\gamma \in \{\alpha, \beta, ..., \eta\}$, $I_\gamma$ is required at a particular stage in the composition and $(O_\alpha \cup O_\beta \cup ... \cup O_\eta) \supseteq O_r$. This problem is referred to as WS composition. The composition thus produces a specification of how to link the available WS to realize the request.

## 3  Interaction Networks

An interaction network constitutes a convenient way to represent a set of interacting WS. It can be an object of study itself, and it can also be used to improve automated WS composition. In this section, we describe what these networks are and how they can be built.

Generally speaking, we define an interaction network as a directed graph whose nodes correspond to interacting objects and links indicate the possibility for the source nodes to act on the target nodes. In our specific case, a node represents a WS, and a link is created from a node $\alpha$ towards a node $\beta$ if and only if for each input parameter in $I_\beta$, a similar output parameter exists in $O_\alpha$. In other words, the link exists if and only if WS $\alpha$ can provide all the information requested to apply WS $\beta$. In Fig. 2, the left side represents a set of WS with their input and output parameters, whereas the right side corresponds to the associated interaction network. Considering WS $\alpha$ and WS $\beta$, all the inputs of $\beta$, $I_\beta = \{f\}$, are included in the outputs of $\alpha$, $O_\alpha = \{d, e, f\}$, i.e. $I_\beta \subseteq O_\alpha$. Hence, $\alpha$ is able to provide all the information needed to interact with $\beta$. Consequently, a link exists between $\alpha$ and $\beta$ in the interaction network. On the contrary, neither $\alpha$ nor $\beta$ ($O_\alpha = \{d, e, f\}$, $O_\beta = \{g, h\}$), provide all the parameters required by $\gamma$ ($I_\gamma = \{d, g\}$), which is why there is no link pointing towards $\gamma$ in the interaction network.

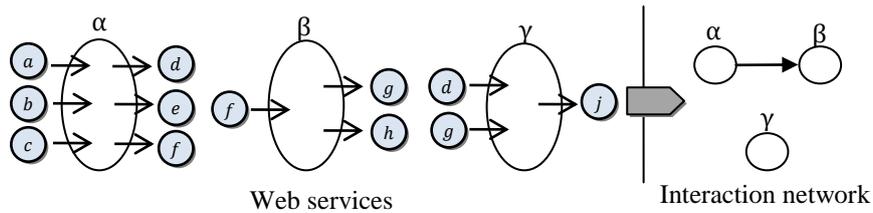

Web services　　　　　　　　　　　Interaction network

**Fig. 2.** Example of a WS interaction network.

An interaction link between two WS therefore represents the possibility of composing them. Determining if two parameters are similar is a complex task which depends on how the notion of similarity is defined. This is implemented under the form of the matching function through the use of similarity metrics.

Parameters similarity is performed on parameter names. A matching function $f$ takes two parameter names $p_1$ and $p_2$, and determines their level of similarity. We use an approximate matching in which two names are considered similar if the value of the similarity function is above some threshold. The key characteristic of the syntactic matching techniques is they interpret the input in function of its sole structure. Indeed, string-based terminological techniques consider a term as a sequence of character. These techniques are typically based on the following intuition: the more similar the strings, the more likely they convey the same information.

We selected three variants of the extensively used edit distance: Levenshtein, Jaro and Jaro-Winkler [6]. The edit distance is based on the number of insertions, deletions, and substitutions of characters required to transform one compared string into the other.

The Levenshtein metric is the basic edit distance function, which assigns a unit cost to all edit operations. For example, the number of operations to transform both strings *kitten* and *sitting* into one another is 3: 1) *k*itten (substitution of *k* with *s*) *s*itten; 2) sitt*e*n (substitution of *e* with *i*) sitt*i*n; 3) sitt*i*n (insertion of *g* at the end) sittin*g*.

The Jaro metric takes into account typical spelling deviations between strings. Consider two strings $s_1$ and $s_2$. A character $a$ in $s_1$ is "in common" with $s_2$ if the same character $a$ appears in about the place in $s_2$. In equation 1, $m$ is the number of matching characters and $t$ is the number of transpositions. A transposition is the operation needed to permute two matching characters if they are not farther than the distance expressed by equation 2.

$$d_j = \frac{1}{3}\left(\frac{m}{|s_1|} + \frac{m}{|s_2|} + \frac{m-t}{m}\right) \quad (1)$$

$$\left\lfloor \frac{\max(|s_1|,|s_2|)}{2} \right\rfloor - 1 \quad (2)$$

The Jaro-Winkler metric, equation 3, is an extension of the Jaro metric. It uses a prefix scale $p$ which gives more favorable ratings to strings that match from the beginning for some prefix length $l$.

$$d_w = d_j + \left(lp(1 - d_j)\right) \quad (3)$$

The metrics score are normalized such that 0 equates to no similarity and 1 is an exact match.

## 4   Network Properties

The degree of a node is the number of links connected to this node. Considered at the level of the whole network, the degree is the basis of a number of measures. The minimum and maximum degrees are the smallest and largest degrees in the whole network, respectively. The average degree is the average of the degrees over all the nodes. The degree correlation reveals the way nodes are related to their neighbors according to their degree. It takes its value between $-1$ (perfectly disassortative) and $+1$ (perfectly assortative). In assortative networks, nodes tend to connect with nodes of similar degree. In disassortative networks, nodes with low degree are more likely connected with highly connected ones [7].

The density of a network is the ratio of the number of existing links to the number of possible links. It ranges from $0$ (no link at all) to $1$ (all possible links exist in the network, i.e. it is completely connected). Density describes the general level of connectedness in a network. A network is complete if all nodes are adjacent to each other. The more nodes are connected, the greater the density [8].

Shortest paths play an important role in the transport and communication within a network. Indeed, the geodesic provides an optimal path way for communication in a network. It is useful to represent all the shortest path lengths of a network as a matrix in which the entry is the length of the geodesic between two distinctive nodes. A measure of the typical separation between two nodes in the network is given by the average shortest path length, also known as average distance. It is defined as the average number of steps along the shortest paths for all possible pairs of nodes [7].

In many real-world networks it is found that if a node $A$ is connected to a node $B$, and $B$ is itself connected to another node $C$, then there is a high probability for $A$ to be also connected to $C$. This property is called transitivity (or clustering) and is formally defined as the triangle density of the network. A triangle is a structure of three completely connected nodes. The transitivity is the ratio of existing to possible triangles in the considered network [9]. Its value ranges from $0$ (the network does not contain any triangle) to $1$ (each link in the network is a part of a triangle). The higher the transitivity is, the more probable it is to observe a link between two nodes possessing a common neighbor.

## 5 Experiments

In those experiments, our goal is twofold. First we want to compare different metrics in order to assess how the links creation is affected by the similarity between the parameters in our interaction network. We would like to identify the best metric in terms of suitability regarding the data features. Second we want to isolate a threshold range within which the matching results are meaningful. By tracking the evolution of the network links, we will be able to categorize the metrics and to determine an acceptable threshold value. We use the previously mentioned complex network properties to monitor this evolution. We start this section by describing our method. We then give the results and their interpretation for each of the topological property mentioned in section 4.

We analyzed the SAWSDL-TC1 collection of WS descriptions [10]. This test collection provides 894 semantic WS descriptions written in SAWSDL, and distributed over 7 thematic domains (education, medical care, food, travel, communication, economy and weapon). It originates in the OWLS-TC2.2 collection, which contains real-world WS descriptions retrieved from public IBM UDDI registries, and semi-automatically transformed from WSDL to OWL-S. This collection was subsequently re-sampled to increase its size, and converted to

SAWSDL. We conducted experiments on the interaction networks extracted from SAWSDL-TC1 using the WS network extractor WS-NEXT [11]. For each metric, the networks are built by varying the threshold from 0 to 1 with a 0.01 step.

Fig. 3 shows the behavior of the average degree versus the threshold for each metric. First, we remark the behavior of the Jaro and the Jaro-Winkler curves are very similar. This is in accordance with the fact the Jaro-Winkler metric is a variation of the Jaro metric, as previously stated. Second, we observe the three curves have a sigmoid shape, i.e. they are divided in three areas: two plateaus separated by a slope. The first plateau corresponds to high average degrees and low threshold values. In this area the metrics find a lot of similarities, allowing many links to be drawn. Then, for small variations of the threshold, the average degree brutally decreases. The second plateau corresponds to average degrees comparable with values obtained for a threshold set at 1, and deserves a particular attention, because this threshold value causes links to appear only in case of exact match. We observe that each curve inflects at a different threshold value. The curves inflects at 0.4, 0.7 and 0.75 for Levenshtein, Jaro and Jaro-Winkler, respectively. Those differences are related to the number of similarities found by the metrics. With a threshold of 0.75, they retrieve 513, 1058 and 1737 similarities respectively.

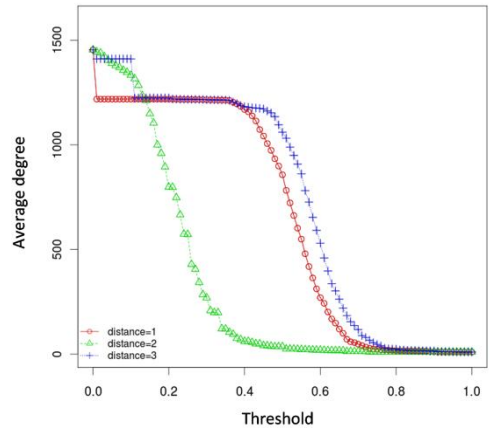

**Fig. 3.** Average degree in function of the metric threshold. Comparative curves of the Levenshtein (green triangles), Jaro (red circles) and Jaro-Winkler (blue crosses) metrics.

To highlight the difference between the curves, we look at their meaningful part, ranging from the inflexion point to the threshold value of 1. We calculated the percentage of average degrees in addition to the average degree obtained with a threshold of 1 for different threshold values. The results are gathered in Table 1. For a threshold of 1, the average degree is 10 and the percentage reference is of course 0%. In the threshold area ranging from the inflexion point to 1, the average degree variation is always above 300%, which seems excessive. Nevertheless, this point needs to be confirmed. Let us assume that above 20% of the minimum average degree, results may be not acceptable (20% corresponding to an average degree of 12). From this postulate, the appropriate threshold is 0.7 for the Levenshtein metric, 0.88 for the Jaro metric. For the Jaro-Winkler metric, the percentage of 17.5 is reached at a threshold of 0.91, then it jumps to 25.4 at the threshold of 0.9. Therefore, we can assume that the threshold range that can be used is [0.7 ; 1] for Levenshtein, [0.88 ; 1] for Jaro and [0.91 ; 1] for Jaro-Winkler.

**Table 1.** Proportional variation in average degree between the networks obtained for some given thresholds and those resulting from the maximal threshold. For each metric, the smaller considered threshold corresponds to the inflexion point.

| Threshold | 0.4 | 0.5 | 0.6 | 0.7 | 0.75 | 0.8 | 0.9 | 1 |
|---|---|---|---|---|---|---|---|---|
| Levenshtein | 510 | 260 | 90 | 20 | 0 | 0 | 0 | 0 |
| Jaro | - | - | - | 370 | 130 | 60 | 10 | 0 |
| Jaro-Winkler | - | - | - | - | 350 | 140 | 50 | 0 |

To go deeper, one has to consider the qualitative aspects of the results. In other words, we would like to know if the additional links are appropriate i.e. if they correspond to parameters similarities having a semantic meaning. To that end, we analyzed the parameters similarities computed by each metric from the 20% threshold values and we estimated the false positives. As we can see in Table 2, the metrics can be ordered according to their score: Jaro returns the least false positives, Levenshtein stands between Jaro and Jaro-Winckler, which retrieves the most false positives. The score of Jaro-Winkler can be explained by analyzing the parameters names. This result is related to the fact this metric favors the existence of a common prefix between two strings. Indeed, in those data, a lot of parameters names belonging to the same domain start with the same beginning. The meaningful part of the parameter stands at the end. As an example, let us mention the two parameter names `Provide MedicalFlightInformation_DesiredDepartureAirport` and `Provide MedicalFlightInformation_DesiredDepartureDateTime`. Those parameters were considered as similar although the end parts have not the same meaning. We find that Levenshtein and Jaro have a very similar behavior concerning the false positives. Indeed, the first false positives that appear are names differing by a very short but very meaningful sequence of characters. As an example, consider: `ProvideMedicalTransportInformation_DesiredDepartureDateTime` and `ProvideNonMedicalTransportInformation_DesiredDepartureDateTime`.
The string `Non` gives a completely different meaning to both parameters, which cannot be detected by the metrics.

**Table 2.** Parameters similarities from the 20% threshold values. 385 similarities are retrieved at the 1 threshold.

| Metric | 20% threshold value | Number of retrieved similarities | Number of false positives | Percentage of false positives |
|---|---|---|---|---|
| Levenshtein | 0.70 | 626 | 127 | 20.3% |
| Jaro | 0.88 | 495 | 53 | 10.7% |
| Jaro-Winkler | 0.91 | 730 | 250 | 34.2% |

To refine our conclusions on the best metric and the most appropriate threshold for each metric, we decided to identify the threshold values leading to false positives. With the Levenshtein, Jaro and Jaro-Winkler metric, we have no false positive at the thresholds of 0.96, 0.98, and 0.99, respectively. Compared to the 385 appropriate similarities retrieved with a threshold of 1, they find 4, 5 and 10 more appropriate similarities, respectively. In Table 3, we gathered the additional similarities retrieved by each metric. At the considered thresholds, it appears that Levenshtein finds some similarities that neither Jaro nor Jaro-Winkler find. Jaro-Winkler retrieves all the similarities found by Jaro and some additional ones. We also analyzed the average degree value at those thresholds. The network extracted with Levensthein does not present an average degree different from the one observed at a threshold of 1. Jaro and Jaro-Winkler networks show an average degree which is 0.52% above the one obtained for a threshold of 1. Hence, if the criterion is to retrieve 0% of false positives, Jaro-Winkler is the most suitable metric.

**Table 3.** Additional appropriate similarities for each metric at the threshold of 0% of false positives.

| Metric Threshold | Similarities |
|---|---|
| Levenshtein 0.96 | `GetPatientMedicalRecords_PatientHealthInsuranceNumber ~ SeePatientMedicalRecords_PatientHealthInsuranceNumber`<br>`_GOVERNMENT-ORGANIZATION ~ _GOVERNMENTORGANIZATION`<br>`_GOVERMENTORGANIZATION ~ _GOVERNMENTORGANIZATION`<br>`_LINGUISTICEXPRESSION ~ _LINGUISTICEXPRESSION1` |
| Jaro 0.98 | `_GOVERNMENT-ORGANIZATION ~ _GOVERNMENTORGANIZATION`<br>`_LINGUISTICEXPRESSION ~ _LINGUISTICEXPRESSION1`<br>`_GEOGRAPHICAL-REGION ~ _GEOGRAPHICAL-REGION1`<br>`_GEOGRAPHICAL-REGION ~ _GEOGRAPHICAL-REGION2`<br>`_GEOPOLITICAL-ENTITY ~ _GEOPOLITICAL-ENTITY1` |
| Jaro-Winkler 0.99 | `_GOVERNMENT-ORGANIZATION ~ _GOVERNMENTORGANIZATION`<br>`_GEOGRAPHICAL-REGION ~ _GEOGRAPHICAL-REGION1`<br>`_GEOGRAPHICAL-REGION ~ _GEOGRAPHICAL-REGION2`<br>`_GEOPOLITICAL-ENTITY ~ _GEOPOLITICAL-ENTITY1`<br>`_LINGUISTICEXPRESSION ~ _LINGUISTICEXPRESSION1`<br>`_SCIENCE-FICTION-NOVEL ~ _SCIENCEFICTIONNOVEL`<br>`_GEOGRAPHICAL-REGION1 ~ _GEOGRAPHICAL-REGION2`<br>`_TIME-MEASURE ~ _TIMEMEASURE`<br>`_LOCATION ~ _LOCATION1`<br>`_LOCATION ~ _LOCATION2` |

The variations observed for the density are very similar to those discussed for the average degree. At the threshold of 0, the density is rather high, with a value of 0.93. Nevertheless, we do not reach a complete network whose density is equal to 1. This is due to the interaction network definition, which implies that for a link to be drawn from a WS to another, all the required parameters must be provided. At the threshold of 1, the density drops to 0.006. At the inflexion points, the density for Levenshtein is 0.038, whereas it is 0.029 for both Jaro and Jaro-Winkler. The variations observed are of the same order of magnitude than those observed for the average degree. For the Levenshtein metric the variation is 533% while for both other metrics it reaches 383%. Considering a density value 20% above the density at the threshold of 1, which is 0.0072, this density is reached at the following thresholds: 0.72 for Levenshtein, 0.89 for Jaro and 0.93 for Jaro-Winkler. The corresponding percentages of false positives are 13.88%, 7.46% and 20.18%. Those values are comparable to the ones obtained for the average degree. Considering the thresholds at which no false positive is retrieved (0.96, 0.98 and 0.99), the corresponding densities are the same that the density at the threshold of 1 for the three metrics. The density is a property which is less sensible to small variations of the number of similarities than the average degree. Hence, it does not allow concluding which metric is the best at those thresholds.

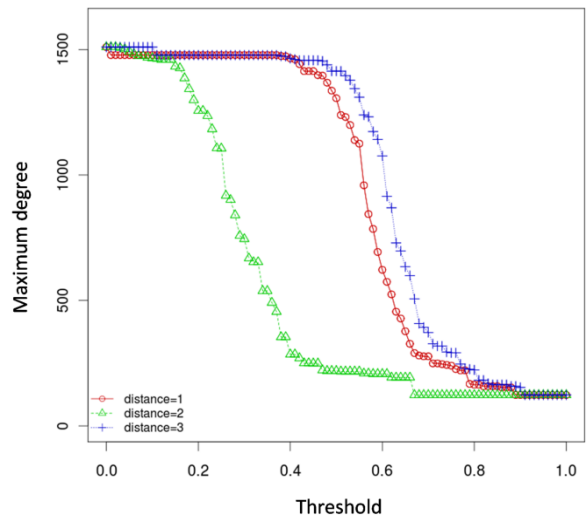

**Fig. 4.** Maximum degree in function of the metric threshold. Comparative curves of the Levenshtein (green triangles), Jaro (red circles) and Jaro-Winkler (blue crosses) metrics.

The maximum degree (cf. Fig. 4) globally follows the same trend than the average degree and the density. At the threshold of 0 and on the first plateau, the maximum degree is around 1510. At the threshold of 1, it falls to 123. Hence, the maximum degree is roughly multiplied by 10. At the inflexion points, the maximum degree is 285, 277 and 291 for Levenshtein, Jaro and Jaro-Winkler respectively. The variations are all of the same order of magnitude and smaller than the variations of the average degree and the density. For Levenshtein, Jaro and Jaro-Winkler the variations values are 131%, 125% and 137% respectively. Considering the maximum degree 20% above 123, which is 148, this value is approached within the threshold ranges [0.66,0.67], [0.88,0.89], [0.90,0.91] for Levenshtein, Jaro and Jaro-Winkler respectively. The corresponding maximum degrees are [193,123] for Levenshtein and [153,123] for both Jaro and Jaro-Winkler. The corresponding percentages of false positives are [28.43%, 26.56%], [10.7%, 7.46%] and [38.5%, 34.24%] . Results are very similar to those obtained for the average degree and the metrics can be ordered the same way. At the thresholds where no false positive is retrieved (0.96, 0.98 and 0.99), the maximum degree is not different from the value obtained with a threshold of 1. This is due to the fact few new similarities are introduced in this case. Hence, no conclusion can be given on which one of the three metric is the best.

As shown in Fig. 5, the curves of the minimum degree are also divided in three areas: one high plateau and one low plateau separated by a slope. A the threshold of 0, the minimum degree is 744. At the threshold of 1, the minimum degree is 0. This value corresponds to isolated nodes in the network. The inflexion points here appear latter: at 0.06 for Levenshtein and at 0.4 for both Jaro and Jaro-Winkler. The corresponding minimum degrees are 86 for Levenshtein and 37 for Jaro and Jaro-Winkler. The thresholds at which the minimum degree starts to be different from 0 are 0.18 for Levenshtein with a value of 3, 0.58 for Jaro with a value of 2, and 0.59 for Jaro-Winkler with a value of 1. The minimum degree is not very sensible to the variations of the number of similarities. Its value starts to increase at a threshold where an important number of false positive have been introduced.

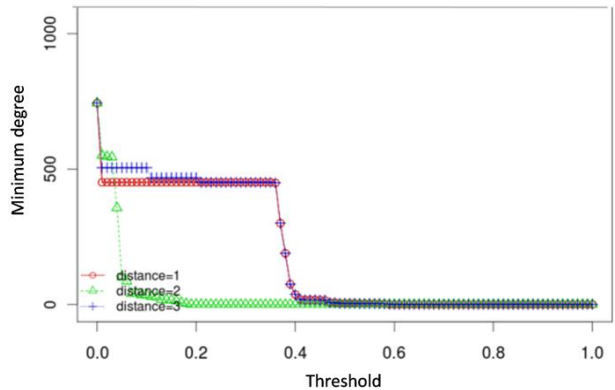

**Fig. 5.** Minimum degree in function of the metric threshold. Comparative curves of the Levenshtein (green triangles), Jaro (red circles) and Jaro-Winkler (blue crosses) metrics.

The transitivity curves (Fig. 6) globally show the same evolution than the ones of the average degree, the maximum degree and the density. The transitivity at the threshold of 0 almost reaches the value of 1. Indeed, the many links allow the existence of numerous triangles. At the threshold of 1, the value falls to 0.032. At the inflexion points, the transitivity values for Levenshtein, Jaro and Jaro-Winkler are 0.17, 0.14 and 0.16 respectively. In comparison with the transitivity at a threshold level of 1, the variations are 431%, 337%, 400%. They are rather high and of the same order than the ones observed for the average degree. Considering the transitivity value 20% above the one at a threshold of 1, which is 0.0384, this value is reached at the threshold of 0.74 for Levenshtein, 0.9 for Jaro and 0.96 for Jaro-Winkler. Those thresholds are very close to the one for which there is no false positive. The corresponding percentages of false positives are 12.54%, 6.76% and 7.26%. Hence, for those threshold values, we can rank Jaro and Jaro-Winkler at the same level, Levensthein being the least performing. Considering the thresholds at which no false positive is retrieved, (0.96, 0.98 and 0.99), the corresponding transitivity are the same than the transitivity at 1. For this reason and by the same way than for the density and the maximum degree, no conclusion can be given on the metrics.

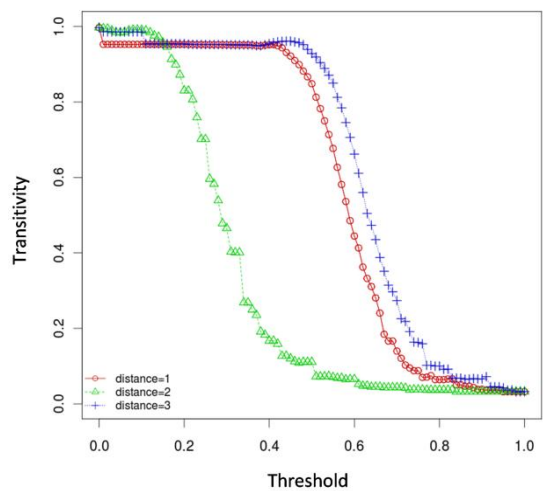

**Fig. 6.** Transitivity in function of the metric threshold. Comparative curves of the Levenshtein (green triangles), Jaro (red circles), and Jaro-Winkler (blue crosses) metrics.

The degree correlation curves are represented in Fig. 7. We can see that the Jaro and the Jaro-Winkler curves are still similar. Nevertheless, the behavior of the three curves is different from what we have observed previously. The degree correlation variations are of lesser magnitude than the variations of the other metrics. For low thresholds, curves start by a stable area in which the degree correlation value is 0. This indicates that no correlation pattern emerges in this area. For high thresholds the curves decrease until they reach a constant value ($-0.246$). This negative value reveals a slight disassortative degree correlation pattern. Between those two extremes, the curves exhibit a maximum value that can be related to the variations of the minimum degree and to the maximum degree. Starting from a threshold value of 1 the degree correlation remains constant until a threshold value of 0.83, 0.90 and 0.94 for Lenvenshtein, Jaro and Jaro-Winkler respectively.

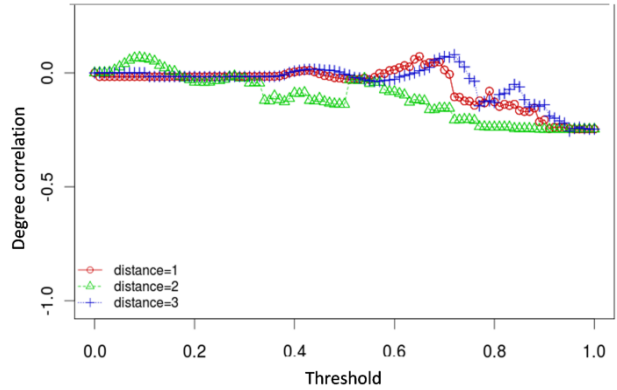

**Fig. 7.** Degree correlation in function of the metric threshold. Comparative curves of the Levenshtein (green triangles), Jaro (red circles) and Jaro-Winkler (blue crosses) metrics.

Fig. 8 shows the variation of the average distance according to the threshold. The three curves follow the same trends and Jaro and Jaro-Winkler are still closely similar. Nevertheless, the curves behavior is different from what we observed for the other properties. For the three metrics, we observe that the average distance globally increases with the threshold until it reaches a maximum value and then start to decrease. The maximum is reached at the thresholds of 0.5 for Levenshtein, 0.78 Jaro and 0.82 Jaro-Winkler. The corresponding average distance values are 3.30, 4.51 and 5.00 respectively. Globally the average distance increases with the threshold. For low threshold values the average distance is around 1 while for the threshold of 1, networks have an average distance of 2.18. Indeed, it makes sense to observe a greater average distance when the network contains less links. This means that almost all the nodes are neighbors of each other. This is in accordance with the results of the density which is not far from the value of 1 for small thresholds. We remark that the curves start to increase as soon as isolated nodes appear. Indeed, the average distance calculation is only performed on interconnected nodes. The thresholds associated to the maximal average distance correspond to the inflexion points in the maximum degree curves. The thresholds for which the average distance stays stable correspond to the thresholds in the maximum degree curves for which the final value of the maximum degree start to be reached. Hence from the observation of the average distance, we can refine the conclusions from the maximum degree curves by saying that the lower limit of acceptable thresholds is 0.75, 0.90 and 0.93 for Levenshtein, Jaro and Jaro-Winkler respectively.

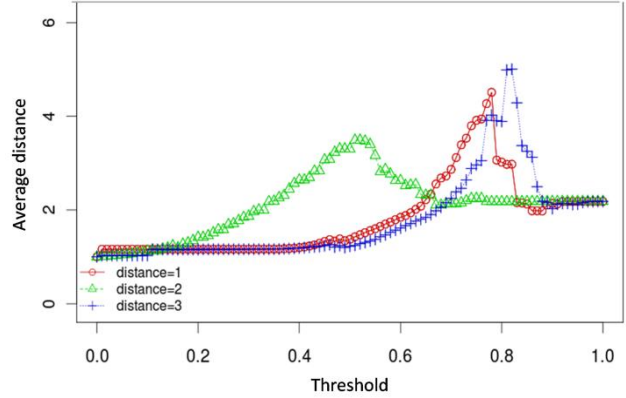

**Fig. 8.** Average distance in function of the metric threshold. Comparative curves of the Levenshtein (green triangles), Jaro (red circles) and Jaro-Winkler (blue crosses) metrics.

## 6 Conclusion

In this work, we studied different metrics used to build WS composition networks. To that end we observed the evolution of some complex network topological properties. Our goal was to determine the most appropriate metric for such an application as well of the most appropriate threshold range to be associated to this metric. We used three well known metrics, namely Levenshtein, Jaro and Jaro-Winkler, especially designed to compute similarity relation between strings. The evolution of the networks from high to low thresholds reflects a growth of the interactions between WS, and hence, of potential compositions. New parameter similarities are revealed, and links are consequently added to the network, along with the threshold increase. If one is interested by a reasonable variation of the topological properties of the network as compared to a threshold value of 1, it seems that the Jaro metric is the most appropriate, as this metric introduces less false positives (inappropriate similarities) than the others. The threshold range that can be associated to each metric is globally [0.7,1], [0.89,1] and [0.91,1] for Levenshtein, Jaro and Jaro-Winkler, respectively. We also examined the behavior of the metrics when no false positive is introduced and new similarities are all semantically meaningful. In this case, Jaro-Winkler gives the best results. Naturally the threshold ranges are lower in this case, and the topological properties are very similar to the ones obtained with a threshold value of 1.

Globally, the use of the metrics to build composition networks is not very satisfying. As the threshold decreases, the false positive rate becomes very quickly prohibitive. This leads us to turn to an alternative approach. It consists in exploiting the latent semantics in parameters name. To extend our work, we plan map the names to ontological concepts with the use of some knowledge bases, such as WordNet [12] or DBPedia [13]. Hence, we could provide a large panel on the studied network properties according to the way similarities are computed to build the networks.